\documentclass[english]{iopart}
\usepackage[T1]{fontenc}
\usepackage[latin9]{inputenc}
\pdfoutput=1
\usepackage{amstext}
\usepackage{amssymb}
\usepackage{graphicx}

\makeatletter
\usepackage{iopams}
\usepackage{setstack}

\makeatother

\usepackage{babel}
\begin{document}
\global\long\def\ad{\hat{a}^{\dagger}}

\global\long\def\aa{\hat{a}}

\global\long\def\bd{\hat{b}^{\dagger}}

\global\long\def\bb{\hat{b}}

\title{Optomechanical circuits for nanomechanical continuous variable quantum
state processing }

\author{Michael Schmidt}

\address{Friedrich-Alexander-Universität Erlangen-Nürnberg, Staudtstr. 7,
D-91058 Erlangen, Germany}

\author{Max Ludwig}

\address{Friedrich-Alexander-Universität Erlangen-Nürnberg, Staudtstr. 7,
D-91058 Erlangen, Germany}

\author{Florian Marquardt}

\address{Friedrich-Alexander-Universität Erlangen-Nürnberg, Staudtstr. 7,
D-91058 Erlangen, Germany}

\address{Max Planck Institute for the Science of Light, Günther-Scharowsky-Straße
1/Bau 24, D-91058 Erlangen, Germany}
\begin{abstract}
We propose and analyze a nanomechanical architecture where light is
used to perform linear quantum operations on a set of many vibrational
modes. Suitable amplitude modulation of a single laser beam is shown
to generate squeezing, entanglement, and state-transfer between modes
that are selected according to their mechanical oscillation frequency.
Current optomechanical devices based on photonic crystals, as well
as other systems with sufficient control over multiple mechanical
modes, may provide a platform for realizing this scheme.
\end{abstract}

\pacs{42.50.Wk, 07.10.Cm, 03.67.Bg, 03.67.Lx}

\maketitle

\section{Introduction}

The field of cavity optomechanics studies the interaction between
light and mechanical motion, with promising prospects in fundamental
tests of quantum physics, ultrasensitive detection, and applications
in quantum information processing (see \cite{Kippenberg2008,Favero2009,MarquardGirvin2009}
for reviews). One particularly promising platform consists of ``optomechanical
crystals'', in which strongly localized optical and vibrational modes
are implemented in a photonic crystal structure \cite{Eichenfield2009}.
So far, several interesting possibilities have been pointed out that
would make use of multi-mode setups designed on this basis. For example,
suitably engineered setups may coherently convert phonons to photons
\cite{Safavi-NaeiniPainter2011} and collective nonlinear dynamics
might be observed in optomechanical arrays \cite{Heinrich2011}. Moreover,
optomechanical systems in general have been demonstrated to furnish
the basic ingredients for writing quantum information from the light
field into the long-lived mechanical modes \cite{Safavi-Naeini2011SlowLight,FioreWang2011,Verhagen2012}.
The recent success in ground state laser-cooling \cite{Teufel2011,chan2011}
has now opened the door to coherent quantum dynamics in optomechanical
systems. 

In this paper, we propose a general scheme for continuous-variable
quantum state processing \cite{Braunstein2005} utilizing the vibrational
modes of such structures. We will show how entanglement and state
transfer operations can be applied selectively to pairs of modes,
by suitable intensity modulation of a single incoming laser beam.
We will discuss limitations for entanglement generation and transfer
fidelity, and show how to pick suitable designs to address these challenges.

\section{Model}

We will first restrict our attention to a single optical mode coupled
to many mechanical modes, such that the following standard optomechanical
Hamiltonian describes the photon field $\hat{a}$, the phonons $\bb_{l}$
of different localized vibrational modes ($l=1,2,\ldots,N$), their
mutual coupling, and the laser drive:

\begin{equation}
\hat{H}=-\hbar\Delta\ad\aa+\sum_{l}\hbar\Omega_{l}\bd_{l}\bb_{l}-\hbar\ad\aa\sum_{l}g_{0}^{(l)}(\bd_{l}+\bb_{l})+\hbar\alpha_{L}(\hat{a}+\hat{a}^{\dagger})+\dots\,.\label{eq:Model}
\end{equation}
\begin{figure}
\includegraphics[width=1\columnwidth]{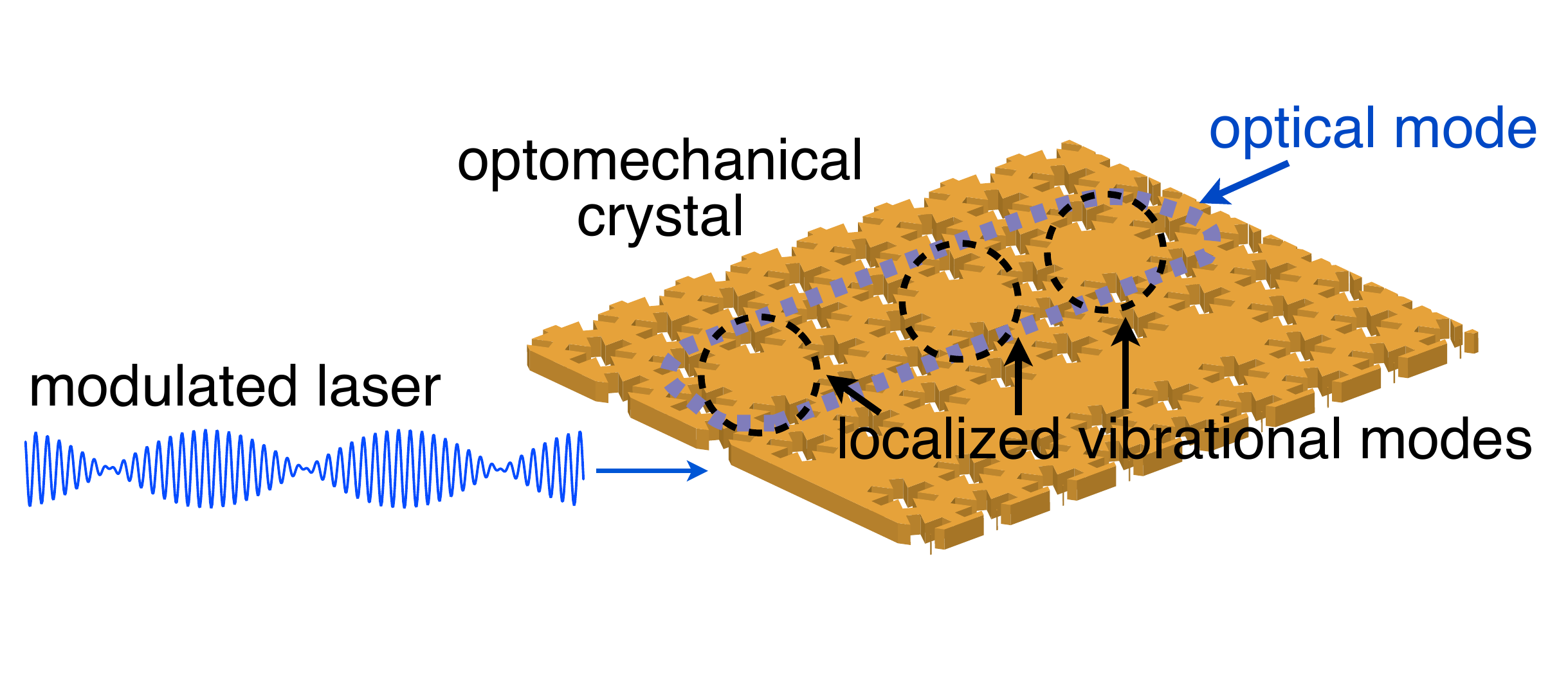}

\caption{\label{SetupFigure}Schematic setup figure illustrating an optomechanical
crystal with localized vibrational modes, coupled by a common optical
mode. The optical mode is driven by an amplitude-modulated laser beam
to engineer frequency-selective entanglement, squeezing and state-transfer
operations of vibrations (see main text).}
\end{figure}
 We are working in a frame rotating at the laser frequency $\omega_{L}$
with detuning $\Delta=\omega_{L}-\omega_{{\rm cav}}$. Here, $\alpha_{L}$
denotes the coupling to the drive, which is proportional to the laser
amplitude. We omitted to explicitly write down the coupling to the
photon and phonon baths, with damping rates $\kappa$ and $\Gamma_{l}$,
respectively, although these will be included in our treatment. The
bare (single-photon) coupling constants $g_{0}^{(l)}$ depend on the
overlap between the optical and mechanical mode functions and are
on the order of $\omega_{{\rm cav}}x_{{\rm ZPF}}/L$, where $x_{{\rm ZPF}}=(\hbar/2m_{l}\Omega_{l})^{1/2}$
is the mechanical zero-point amplitude. In photonic crystals the effective
cavity length $L$ reaches down to wavelength dimensions (see Figure~\ref{SetupFigure}
for the illustration of a setup).

After application of the standard procedure of splitting off the coherent
optical amplitude induced by the laser, $\aa=\alpha+\delta\aa$, and
omitting terms quadratic in $\delta\aa$ (valid for strong drive)
to the Hamiltonian, we recover the linearized optomechanical coupling,

\begin{equation}
\hat{H}_{{\rm int}}=-\hbar(\delta\aa+\delta\ad)\sum_{l}g_{l}(\bd_{l}+\bb_{l})\,=-\hbar(\delta\aa+\delta\ad)\hat{Y}.
\end{equation}
Here the dressed couplings $g_{l}=g_{0}^{(l)}\alpha$ can be tuned
via the laser intensity, i.e. the circulating photon number: $|\alpha|=\sqrt{\bar{n}_{{\rm phot}}}$
($\alpha\in\mathbb{R}$ without loss of generality). We can now eliminate
the driven cavity field (noting that $\delta\aa$ is in the ground
state) by second-order perturbation theory: At large detuning $\left|\Delta\right|\gg\Omega_{l}$
the energy scales for phonons and photons separate, and we retain
a fully coherent, light-induced mechanical coupling between all the
vibrational modes,

\begin{equation}
\hat{H}_{{\rm int}}^{{\rm eff}}=\hbar\frac{\hat{Y}^{2}}{\Delta}=2\hbar\sum_{l,k}J_{lk}(t)\hat{X}_{l}\hat{X}_{k}\,,\label{eq:coupling}
\end{equation}
where $\hat{X}_{l}\equiv\bb_{l}+\bd_{l}$ is the mechanical displacement
in units of $x_{{\rm ZPF}}$. For this to be valid we have to fulfill
$\kappa\ll|\Delta|$, which prevents unwanted transitions. Eq.~(\ref{eq:coupling})
may be viewed as a ``collective optical spring'' effect, coupling
all the mechanical displacements. The couplings $J_{lk}(t)=g_{l}g_{k}/2\Delta$
can be changed in-situ and in a time-dependent manner via the laser
intensity or the detuning. In the numerical simulations we take $g_{0}^{(l)}=g_{0}x_{\text{zpf}}^{(l)}/x_{\text{zpf}}^{(1)}=g_{0}\sqrt{\Omega_{1}/\Omega_{l}}$
and assume all modes to have equal masses. This feature will be crucial
for our approach described below. Note that if multiple optical modes
are driven, the corresponding coupling constants will add. 

In general, the couplings in Eq.~(\ref{eq:coupling}) will induce
quantum state transfer between mutually resonant mechanical modes,
and entanglement at low temperatures (usually with the help of optomechanical
laser cooling). The generation of entanglement between two mechanical
modes based on the bilinear interaction, Eq. (\ref{eq:coupling}),
has been analyzed in \cite{PinardDantanVitaliArcizetBriantHeidmann2005,Ebhardt2008,Bhattacharya2008,Hartmann2008,Hammerer2009,Ludwig2010,Akram2011}
or in \cite{Vitali2007_2,Paternostro2007,Abdi2011,Akram2011} for
entanglement between the light field and the phononic mode. However,
these schemes are not easily scalable to many mechanical modes, mainly
because it is not possible to address the modes that are to be entangled.
Moreover, these schemes produce an entangled steady-state that is
sensitive to thermal fluctuations. We will describe a different scheme
that allows mode selective entanglement and is particularly suited
for muti-mode setups. Its robustness against thermal influences is
enhanced, since it employs parametric instabilities for entanglement
generation.

\section{General scheme}

In contrast to the schemes mentioned above, we have in mind a multi-mode
situation for continuous variable quantum information processing and
are interested in an efficient approach to selectively couple arbitrary
pairs of modes, both for entanglement and state transfer. There are
several desiderata to address for a suitable optomechanical architecture
of that style: one should be able to (i)~switch couplings in time;
(ii)~easily select pairs for operations; (iii)~ get by with only
one laser (or a limited number); (iv) achieve large enough operation
speeds to beat decoherence; (v)~scale to a reasonably large number
of modes.

\subsection{Frequency selective operations}

Static couplings as in Eq.~(\ref{eq:coupling}) could be used for
selective pairwise operations if one were able to shift locally the
mechanical frequency to bring the two respective modes into resonance.
In principle, this is achievable via the optical spring effect, but
would require local addressing with independent laser beams. This
could prove challenging in a micron-scale photonic crystal, severely
hampering scalability.

Instead, we propose to employ frequency-selective operations, by modulating
the laser intensity (and thus $J$) in time. Entanglement generation
by parametric driving has been analyzed recently in various contexts,
including superconducting circuits \cite{Tian2008}, trapped ions
\cite{Serafini2009NJP}\textbf{,} general studies of entanglement
in harmonic oscillators \cite{Galve2009,Bastidas2010,Galve2010},
optomechanical state transfer and entanglement between the motion
of a trapped atom and a mechanical oscillator \cite{Wallquist2010}
and entanglement between mechanical and radiation modes \cite{Mari2011}.
Parametric driving can also lead to mechanical squeezing in optomechanical
systems \cite{Mari2009}.\textbf{ }

Let us now consider two modes with coupling $2\hbar J(t)(\hat{X}_{1}+\hat{X}_{2})^{2}$.
Note that for the purposes of our discussion we set $J_{lk}=J$ for
sake of simplicity. The results mentioned below, however, remain valid
for the general case of unequal couplings, which is also used for
the numerical simulations. The time-dependent coupling is achieved
by modulating the laser intensity at a frequency $\omega$ of the
order of the mechanical frequencies. For $\omega\ll|\Delta|$ the
circulating photon number follows adiabatically $|\alpha(t)|^{2}=|\alpha_{{\rm max}}|^{2}\cos^{2}(\omega t)$,
and we have $J(t)=J\cos^{2}(\omega t)=J[1+\cos(2\omega t)]/2$. The
resulting time-dependent light-induced mechanical coupling can be
broken down into several contributions, whose relative importance
will be determined by the drive frequency $\omega$. The static terms,
$\hbar J(\hat{X}_{1}+\hat{X}_{2})^{2}$, shift the oscillator frequencies
by $\delta\Omega_{j}=2J$, and give rise to an off-resonant coupling
 that is ineffective for $|\Omega_{1}-\Omega_{2}|\gg J$, but   gains
influence for $\text{|}\Omega_{1}-\Omega_{2}|\lesssim J$ . In a realistic
setup this $\hat{X}_{1}\hat{X}_{2}$ interaction might be enhanced
due to intrinsically present phonon tunneling between distinct vibrational
modes, which could easily be included into the analysis, since it
is of the same structure. Moreover the static terms contain single
mode squeezing terms, $\hat{b}_{l}^{\dagger}\hat{b}_{l}^{\dagger}+\hat{b}_{l}\hat{b}_{l}$
that are always off-resonant and of negligible influence . For the
oscillating terms 

\begin{equation}
\hbar J(e^{2i\omega t}+e^{-2i\omega t})(\bb_{1}+\bb_{1}^{\dagger})(\bb_{2}+\bb_{2}^{\dagger}),\label{eq:timeDependentPart}
\end{equation}
there are three important cases. A mechanical beam-splitter (state-transfer)
interaction is selected for a laser drive modulation frequency $\omega=(\Omega_{1}-\Omega_{2})/2$.
In the interaction picture with respect to $\Omega_{1}$ and $\Omega_{2}$,
the resonant part of the full Hamiltonian reads 
\[
\hat{H}_{{\rm b.s.}}=\hbar J(\bb_{2}^{\dagger}\bb_{1}+\bb_{1}^{\dagger}\bb_{2}).
\]
In contrast, for $\omega=(\Omega_{1}+\Omega_{2})/2$, we obtain from
equation (\ref{eq:timeDependentPart}) a two-mode squeezing (non-degenerate
parametric amplifier) Hamiltonian, 
\[
\hat{H}_{{\rm ent}}=\hbar J(\bb_{1}\bb_{2}+\bb_{1}^{\dagger}\bb_{2}^{\dagger}),
\]
which can lead to efficient entanglement between the modes. Finally,
$\omega=\Omega_{j}$ selects the squeezing interaction for a given
mode, 
\[
\hat{H}_{sq}=\hbar(J/2)(\bb_{j}^{2}+\bb_{j}^{\dagger2}),
\]
out of equation (\ref{eq:timeDependentPart}). These laser-tunable,
frequency-selective mechanical interactions are the basic ingredients
for the architecture that we will develop and analyze here. Furthermore,
combinations of these Hamiltonians can be constructed, when the laser
intensity is modulated with multiple frequencies. Note that one has
to keep in mind that the modulation generates also a time dependent
radiation pressure force, $\hbar g_{0}|\alpha_{{\rm max}}|^{2}\cos^{2}(\omega t)/x_{{\rm ZPF}}$,
that leads to a coherent driving of each of the mechanical modes for
$\omega\approx\Omega_{l}/2$. For the parameters we choose here, these
processes are off-resonant, though. 
\begin{figure}
\includegraphics[width=1\columnwidth]{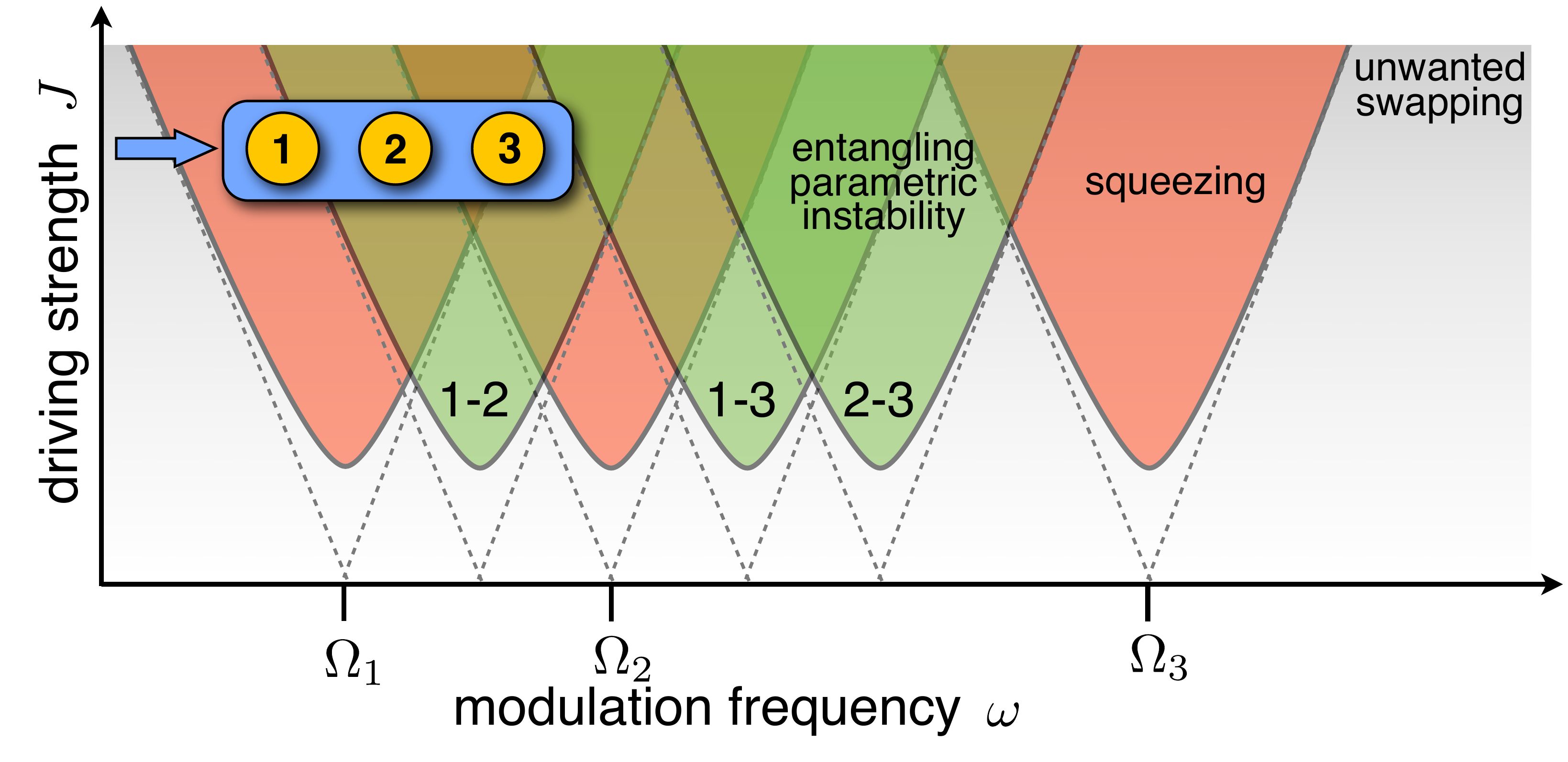}

\caption{\label{SchematicThreeModes}Schematic diagram of the resonance regions
for squeezing and entanglement operations between vibrational modes
($\Omega_{1}$, $\Omega_{2}$ and $\Omega_{3}$), induced by harmonically
modulating the laser intensity at frequency $\omega$. For too weak
laser intensity (small $J$), dissipation prevents parametric resonances
(dashed lines indicate resonance regions without dissipation). With
increasing light-induced coupling $J$ dissipation is overcome and
regions of parametric resonance for entangling and squeezing operations
emerge. By adjusting the modulation frequency $\omega$, these operations
can be addressed selectively. When $J$ is further increased, different
regions start to overlap and selectivity is lost. At large J unwanted
swapping processes occur, leading to an off-resonant coupling between
the modes.}
\end{figure}

\subsection{Limitations}

\begin{figure}
\includegraphics[width=1\columnwidth]{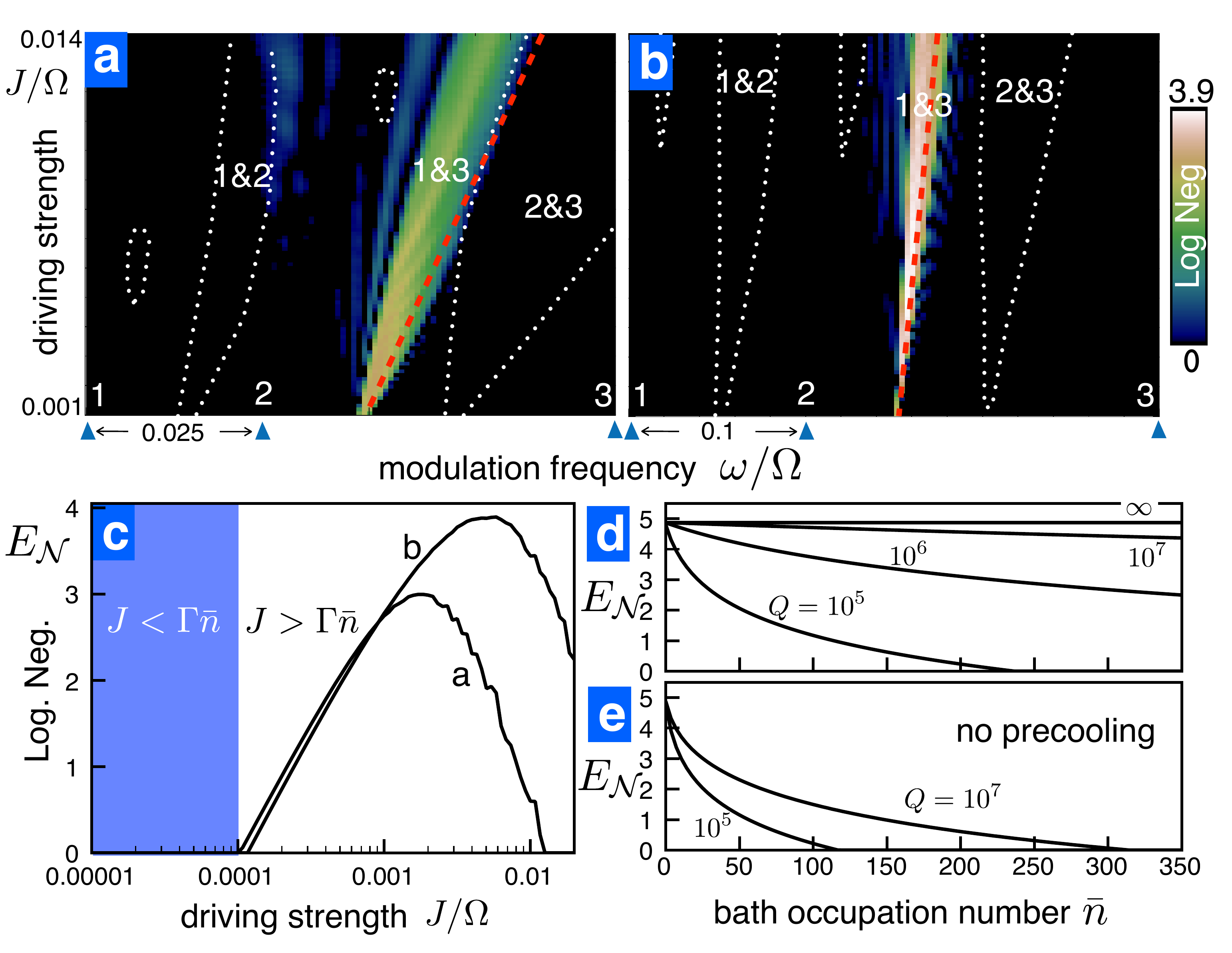}

\caption{\label{EntanglementSimulation}Selective entanglement between two
out of three vibrational modes, produced by an amplitude-modulated
laser beam driving a common optical mode, as sketched in Figure \ref{SchematicThreeModes}.
All modes are coupled to a thermal bath and assumed to be cooled to
their ground state initially. (a,b) Entanglement measure (the logarithmic
negativity) between modes 1 and 3, as a function of the driving strength
$J$ and the modulation frequency $\omega$, at time $t=5.6/J$. {[}Contour
lines: entanglement between 1\&2 or 2\&3 larger than 0.5, blue arrows
at $\Omega_{j}$, red line: resonance position as derived in rotating
wave approximation{]}. In (a) the vibrational modes are spaced densely
$[\Omega_{i}=(1,1.025,1.075)\Omega]$, leading to overlapping resonance
regions (as in Figure~\ref{SchematicThreeModes}). (b) Larger frequency
spacing increases the selectivity significantly {[}$\Omega_{i}=(1,1.1,1.3)\Omega${]}.
(c) Dependence on driving strength, i.e. cut along dashed red lines
in (a,b), see main text. (d) Dependence on temperature and mechanical
quality factor $Q$. (e) Same as (d) but starting from thermal equilibrium.
{[}Parameters: $\bar{n}=100$, $Q=10^{6}$ (a,b,c), $\Omega_{i}=(1,1.1,1.3)\Omega$
(b,d,e), $J/\Omega=0.003$ (d,e), $\Delta=10\Omega,$ $\kappa=\Omega/8$
(a-e){]}}
\end{figure}
We now address the constraining factors for the operation fidelity,
both for the two-mode and the multi-mode case. At higher drive powers
(as needed for fast operations), the frequency-time uncertainty implies
that the different processes need not to be resonant exactly, with
an allowable spread $\left|\delta\omega\right|\lesssim J$. The parametric
instabilities occur for $|\omega-(\Omega_{i}+\Omega_{j})/2|<J$. Once
these intervals start to overlap, process selectivity is lost and
the fidelity suffers. At low operation speeds quantum dissipation
and thermal fluctuations will limit the fidelity. This is the essential
problem faced by a multi-mode setup, and we will discuss possible
remedies further below. The schematic situation for the example of
three modes is illustrated in Figure~\ref{SchematicThreeModes}. 

In order to analyze decoherence and dissipation, we employ a Lindblad
master equation to evolve the joint state of the mechanical modes.
The evolution of any expectation value can be derived from the master
equation and is governed by: 

\begin{eqnarray*}
\frac{d}{dt}\left\langle \hat{A}\right\rangle  & = & \frac{1}{i\hbar}\left\langle [\hat{A},\hat{H}]\right\rangle +\sum_{j}(\bar{n}_{j}+1)\Gamma_{j}\left\langle \mathcal{R}[\hat{b}_{j}^{\dagger}]\hat{A}\right\rangle \\
 &  & +\sum_{j}\bar{n}_{j}\Gamma_{j}\left\langle \mathcal{R}[\hat{b}_{j}]\hat{A}\right\rangle .
\end{eqnarray*}
Here $\hat{H}$ describes the vibrational modes and already contains
the effective interaction (\ref{eq:coupling}), with modulated time-dependent
couplings. $\Gamma_{j}$ are the damping rates of the vibrational
modes, and $\bar{n}_{j}$ their equilibrium occupations at the bulk
temperature. Note that we effectively added the light-induced decoherence
rate $\Gamma_{\text{opt}}^{\varphi}\approx g_{0}^{2}\alpha^{2}\kappa/\Delta^{2}=2J(\kappa/\Delta)$
to the intrinsic rate $\Gamma\bar{n}$ \cite{Marquardt2007QuantumCooling}.
$\Gamma_{\text{opt}}^{\varphi}$ is suppressed by a factor $\kappa/|\Delta|$
and can be arbitrarily small for larger detuning (at the expense of
higher photon number $\alpha^{2}$ to keep the same $J$). In the
numerical simulation we use $J$ and $\Delta$ as independent parameters.
Note, however, that the sign of the detuning affects the sign of $J.$
The dissipative term in the first line describes damping (spontaneous
and induced emission), and the second line refers to absorption of
thermal fluctuations. The relaxation superoperators are defined by
$\mathcal{R}[\hat{b}^{\dagger}]\hat{A}=\hat{b}^{\dagger}\hat{A}\hat{b}-\hat{b}^{\dagger}\hat{b}\hat{A}/2-\hat{A}\hat{b}^{\dagger}\hat{b}/2$
(in contrast to the equation for $\hat{\rho}$). For the quadratic
Hamiltonian studied here, the equations for the correlators remain
closed and are sufficient to describe the Gaussian states produced
in the evolution. 

We evaluate the logarithmic negativity

\begin{equation}
E_{\mathcal{N}}(\hat{\rho})={\rm log}_{2}\left\Vert \hat{\rho}_{AB}^{T_{A}}\right\Vert _{1}
\end{equation}
as a measure of entanglement for any two given modes ($A$ and $B$),
where $\hat{\rho}_{AB}$ is the state of these two modes, and the
partial transpose $T_{A}$ acts on $A$ only. For Gaussian states,
$E_{\mathcal{N}}$ can be calculated via the symplectic eigenvalues
of the position-momentum covariance matrix \cite{Vidal2002}. 

In Figure~\ref{EntanglementSimulation} we show the simulation results
for entangling two out of three vibrational modes. The entanglement
saturates at later times, while the phonon number grows exponentially.
We plot the results at the fixed time $t=5.6/J$, since there the
logarithmic negativity has already saturated. One clearly sees the
features predicted above (Figure \ref{SchematicThreeModes}), i.e.
the unwanted overlap between entanglement processes at higher driving
strengths (Figure~\ref{EntanglementSimulation}a,b). Increasing the
vibrational frequency spacing suppresses these unwanted effects (Figure~\ref{EntanglementSimulation}b).
In Figure~\ref{EntanglementSimulation}c the threshold $J=\Gamma\bar{n}$
for entanglement generation at finite temperature is evident, as is
the loss of entanglement at large $J$. Finally, Figure~\ref{EntanglementSimulation}d,e
shows the dependence on temperature and mechanical quality factor
indicating that this scheme should be feasible for realistic experimental
parameters (see below).

\subsection{Larger arrays}

Having evenly spaced mechanical frequencies is impossible, because
the state transfers between adjacent modes would all be addressed
at the same modulation frequency. Any simple layout that offers selectivity
and avoids resonance overlap seems to require a frequency interval
that grows exponentially with the number $N$ of modes. Hence, another
approach is needed for large $N$. 

\begin{figure}
\includegraphics[width=1\columnwidth]{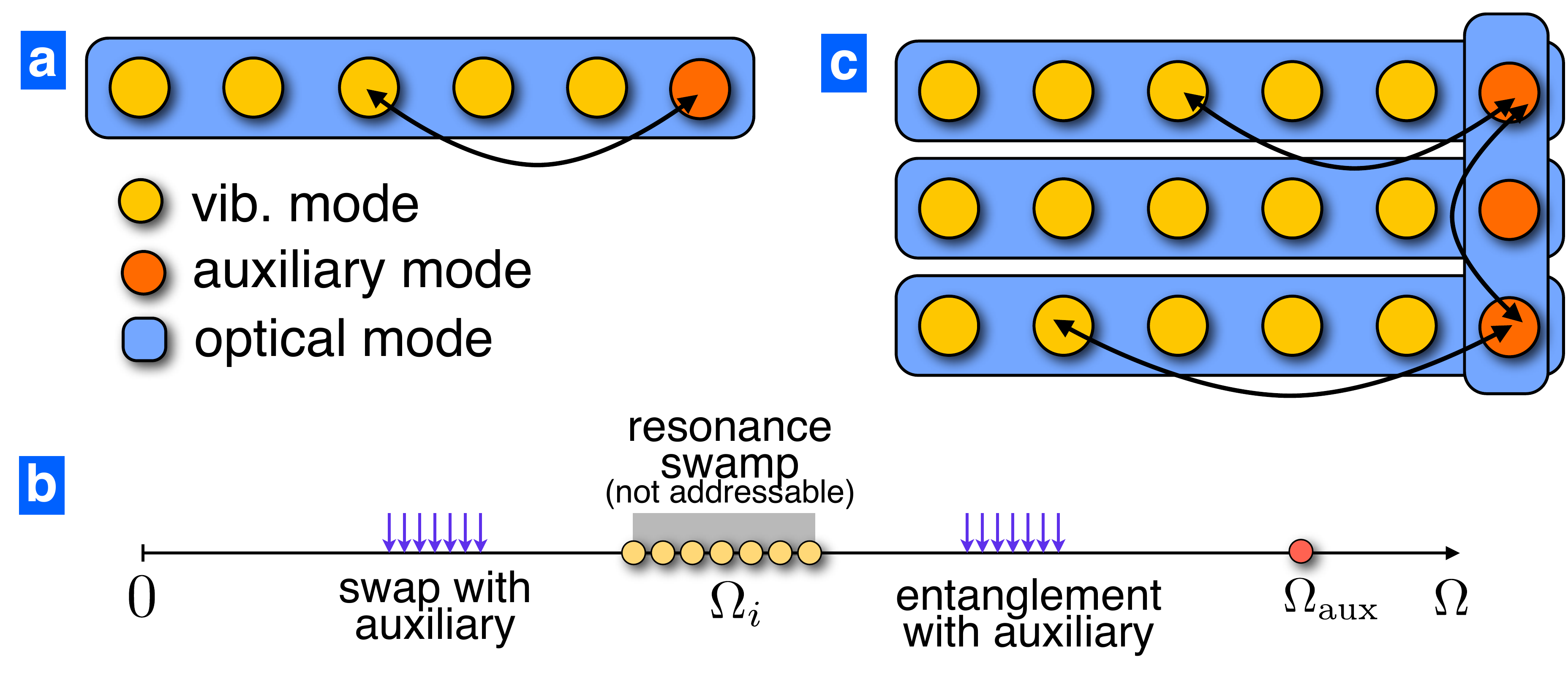}

\caption{\label{LayoutManyModes}(a) Layout for an optomechanical array with
an auxiliary vibrational mode, circumventing the problem of frequency
crowding (see main text). (b) Corresponding mechanical frequency spectrum.
The laser modulation frequency has to lie within certain intervals
to select entanglement or state-transfer operations between the ``memory
modes'' and the auxiliary mode (arrows). (c) Extension to 2D block
consisting of three arrays with operations ``around the corner''.}
\end{figure}

\begin{figure}
\includegraphics[width=1\columnwidth]{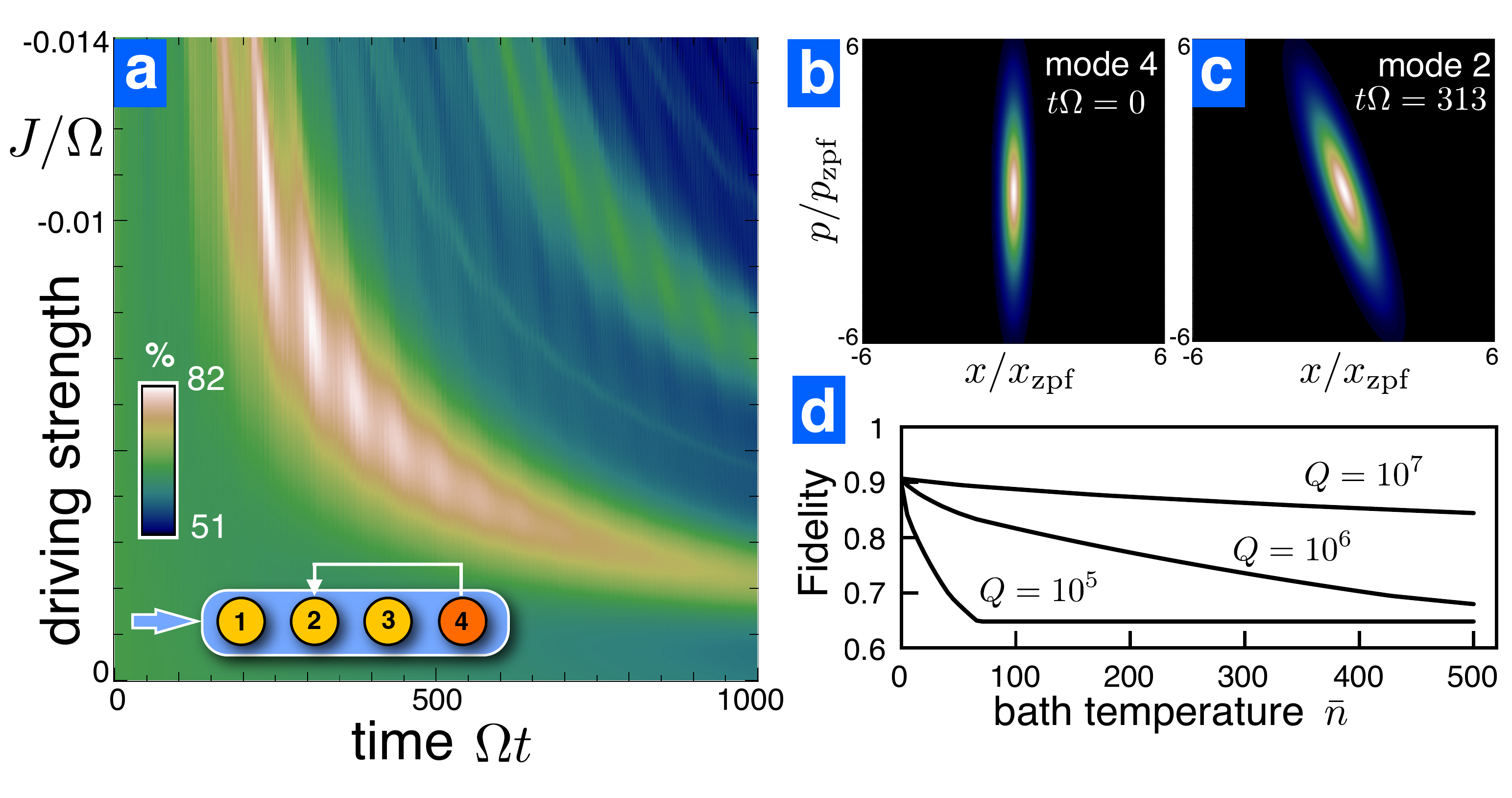}

\caption{\label{StateTransfer} Transfer of a squeezed state ($\Delta x^{2}=e^{-2}x_{{\rm ZPF}}^{2}$)
from the auxiliary to the second memory mode (all in their ground
state due to prior side-band cooling). (a) Dependence of the transfer
fidelity on driving strength (negative J due to red detuning $\Delta<0$)
and pulse time (swap pulse ideally at $|J|t=\pi/2$). For long pulse
times dissipation hampers the transfer, for large driving strengths
resonances overlap, medium times are optimal. (b) Wigner density of
the initial squeezed state in the auxiliary mode, and (c) after the
transfer. Plotted Wigner density maximizes the transfer fidelity for
fixed $J=-0.007\Omega$. (d) Maximization of transfer fidelity over
parameter ranges used in (a). {[}Parameters: $\Omega_{j}=(1.0,1.1,1.2,1.8)\Omega$,
$\bar{n}=100$, $Q=10^{6}$, $\Delta=-20\Omega$, $\kappa=\Omega/8$,
$J/\Omega=-0.007$ (b,c){]} }
\end{figure}

The scheme (Figure~\ref{LayoutManyModes}) that solves this challenge
involves an auxiliary mode at $\Omega_{{\rm aux}}$, removed in frequency
from the array of evenly spaced ``memory'' modes in $[\Omega_{{\rm min}},\,\Omega_{{\rm max}}]$.
All operations will take place between a selected memory mode and
the auxiliary mode. Then, the state transfer resonances are in the
band $[(\Omega_{{\rm aux}}-\Omega_{{\rm max}})/2,(\Omega_{{\rm aux}}-\Omega_{{\rm min}})/2]$,
and entanglement is addressed within $[(\Omega_{{\rm min}}+\Omega_{{\rm aux}})/2,(\Omega_{{\rm max}}+\Omega_{{\rm aux}})/2]$.
To make this work, one needs to fulfill the mild constraint $2\Omega_{{\rm max}}-\Omega_{{\rm min}}<\Omega_{{\rm aux}}<2\Omega_{{\rm min}},$
where the upper limit prevents unintended driving of the modes caused
by the modulated radiation pressure force (see above). Starting with
an arbitrary multimode state, state transfer between two memory modes
is performed in three steps (swapping $1\text{\textendash{\rm aux}}$,
${\rm aux}\text{\textendash2}$ and ${\rm aux}-1$), as is entanglement
(swap $1\text{\textendash{\rm aux}}$, entangle ${\rm aux}\text{\textendash2}$
and swap ${\rm aux}\text{\textendash1}$). Note that this overhead
does not grow with the number of memory modes. Figure~\ref{StateTransfer}
shows the transfer of a squeezed state from the auxiliary mode to
a memory mode. 

Several such arrays could be connected in a 2D scheme by linking their
auxiliaries with an optical mode (Figure~\ref{LayoutManyModes}c).
The spectrum of the auxiliaries can be chosen as in the introductory
example (Figure \ref{SchematicThreeModes}), which is allowed by the
above constraint and enables selective operations between the auxiliaries.
State transfer between memory modes of distinct arrays is possible
in five steps (swap ${\rm mem}_{1}\text{\textendash}{\rm aux}_{1}$,
swap ${\rm mem}_{2}\text{\textendash}{\rm aux}_{2}$ swap ${\rm aux}\text{\textendash}{\rm aux}$,
swap ${\rm aux}_{1}\text{\textendash}{\rm mem}_{1}$, swap ${\rm aux}_{2}\text{\textendash}{\rm mem}_{2}$)
as well as entanglement (swap ${\rm mem}_{1}\text{\textendash}{\rm aux}_{1}$,
swap ${\rm mem}_{2}\text{\textendash}{\rm aux}_{2}$, entangle ${\rm aux}\text{\textendash}{\rm aux}$,
swap ${\rm aux}_{1}\text{\textendash}{\rm mem}_{1}$, swap ${\rm aux}_{2}\text{\textendash}{\rm mem}_{2}$).
Note that the transfer from the memory modes to the auxiliaries can
be done in parallel, hence the scheme can effectively be completed
in three steps and there is no time delay compared to the single array
with auxiliary. Moreover one might employ more sophisticated transfer
schemes \cite{Bergmann1998,Wang2012,Tian2012}, to improve the transfer
fidelities. Two blocks can be connected by introducing a higher order
auxiliary that couples to both auxiliary arrays. State transfer can
then be performed between auxiliary modes from different blocks. In
principle many blocks can be connected, when their auxiliary modes
are stringed together in a chain via higher order auxiliaries.

\section{Implementation}

Regarding the experimental implementation, in principle any optomechanical
system with long-lived mechanical modes can be used. One promising
platform are ``optomechanical crystals'', as introduced by Painter
et al. \cite{Eichenfield2009} that feature vibrational defect cavities
in the GHz regime with experimentally accessible $Q\sim10^{6}$ \cite{Chan2012}.
These would be very well suited for the scheme presented here, due
to their design flexibility, particularly of two dimensional structures,
and the all-integrated approach, as well as the very large optomechanical
coupling strength. Given the currently achieved coupling strength
\cite{chan2011,Chan2012} $g_{0}/2\pi\sim1{\rm MHz}$, a detuning
of $\Delta/\Omega=10$ and around $2000$ cavity photons (reached
in recent experiments), we can estimate the induced coupling to approach
the damping rate, $J\sim\Gamma$. This corresponds to the threshold
for coherent operations, provided one were to cool down the bath to
$k_{B}T_{{\rm bath}}<\hbar\Omega$. This is in principle possible
(at $20{\rm mK}$), but will likely run into practical difficulties
due to the re-heating of the structure via spurious photon absorption
or other effects. At finite bath temperatures corresponding to a thermal
occupation $\bar{n}\sim k_{B}T_{{\rm bath}}/\hbar\Omega$, the light
intensity must be increased by a factor $\bar{n}$ towards $J\gg\Gamma\bar{n}$.
Initially, the vibrational ground state would be prepared via laser-cooling,
as demonstrated in \cite{chan2011}.

In these devices, localized vibrational and optical modes can be produced
at engineered defects in a periodic array of holes cut into a free-standing
substrate. Adjacent optical and vibrational modes are coupled via
tunneling. The typical photon tunnel coupling for modes spaced apart
by several lattice constants is \cite{Heinrich2011} in the range
of several ${\rm THz}$. Thus, hybridized optical modes will form,
one of which can be selected via the laser driving frequency while
the others remain idle. The vibrational modes' frequencies can be
different or equal, in which case delocalized hybridized mechanical
modes are produced. A 'snowflake' crystal made of connected triangles
(honeycomb lattice) supports wave guides (line defects) and localized
defect modes with optomechanical interaction \cite{Safavi-NaeiniPainter2011}.
Placing point defects (heavier triangles/thicker bridges) in the structure,
a tight binding analysis indicates that the mechanical frequency spectrum
(Figure~\ref{LayoutManyModes}b) can be generated. 

Utilizing the phononic modes of an optomechanical system for processing
continuous variable quantum information has a number of desirable
aspects in comparison to other systems like optical modes or cold
atomic vapors. First, the phononic modes can be integrated on a chip
and the devices are thus naturally scalable. The Gaussian operations
discussed above can be applied easily and the decoherence times are
already reasonable, although admittedly worse than for cold atomic
vapors.

We briefly mention another option for improving the fidelity: Optimal
control techniques \cite{Galve2009} could be employed to numerically
optimize the pulse shape $J(t)$.

Finally, an essential ingredient will be read-out. We have pointed
out \cite{Clerk2008} how to produce a quantum-non-demolition read-out
of the mechanical quadratures in an optomechanical setup. A laser
beam (detuning $\Delta=0$) is amplitude-modulated at the mechanical
frequency $\Omega_{j}$ of one of the modes. The reflected light carries
information only about one quadrature $e^{i\varphi}\hat{b}_{j}+e^{-i\varphi}\hat{b}_{j}^{\dagger}$.
Its phase $\varphi$ is selected by the phase of the amplitude-modulation,
while the measurement back-action perturbs solely the other quadrature.
Different modes can be read out simultaneously, and the covariance
matrix may thus be obtained in repeated experimental runs. Taking
measurement statistics for continuously varied quadrature phases would
also allow to do full quantum-state tomography, and thereby ultimately
process tomography. Alternatively, short pulses may be used for readout
(and manipulation) \cite{Vanner2010}.

\section{Conclusions}

The scheme described here would enable coherent scalable nanomechanical
state processing in optomechanical arrays. It can form the basis for
generating arbitrary entangled mechanical Gaussian multi-mode states
like continuous variable cluster states \cite{vanLoock2007}. An interesting
application would be to investigate the decoherence of such states
due to the correlated quantum noise acting on the nanomechanical modes.
Moreover, recent experiments have shown in principle how arbitrary
states can be written from the light field into the mechanics \cite{Safavi-Naeini2011SlowLight,FioreWang2011,Verhagen2012}.
These could then be manipulated by the interactions described here.
Alternatively, for very strong coupling $g_{0}>\kappa$, non-Gaussian
mechanical states \cite{Ludwig2008,Nunnenkamp2011,Qian2011} could
be produced, and the induced nonlinear interactions (see e.g. \cite{Ludwig2012,Stannigel2012})
could potentially open the door to universal quantum computation with
continuous variables \cite{Braunstein2005} in these systems. 

\ack We acknowledge the ITN Cavity Quantum Optomechanics, an ERC
Starting Grant, the DFG Emmy-Noether program and DARPA ORCHID for
funding.

\section*{References}

\bibliographystyle{unsrt}
\bibliography{OMC_Article_Literature_NJP}

\end{document}